\documentclass[preprint,10pt]{elsarticle}

\usepackage[utf8]{inputenc}
\usepackage[english]{babel}
\usepackage{amssymb}
\usepackage[utf8]{inputenc}
\usepackage[english]{babel}
\usepackage{natbib}
\usepackage{hyperref}
\usepackage{amsmath}
\usepackage{longtable}
\usepackage{geometry}
\usepackage{longtable}

\biboptions{sort&compress}
\journal{Nuclear Physics A}

\begin{document}

\begin{frontmatter}



\title{Cluster radioactivity in trans-lead region: A systematic study with modified empirical formulas}


\author[a,b,c]{A. Jain}\author[d]{P. K. Sharma}\author[a]{S. K. Jain}\author[e]{J. K. Deegwal}\author[c,f]{G. Saxena}
\address[a]{Department of Physics, School of Basic Sciences, Manipal University Jaipur, Jaipur-303007, India}
\address[b]{Department of Physics, S. S. Jain Subodh P.G.(Autonomous) College, Jaipur-302004, India}
\address[c]{Department of Physics (H\&S), Govt. Women Engineering College, Ajmer-305002, India}
\address[d]{Govt. Polytechnic College, Rajsamand-313324, India}
\address[e]{Govt. Women Engineering College, Ajmer-305002, India}
\address[f]{Department of Physics, Faculty of Science, University of Zagreb, Bijeni$\breve{c}$ka c. 32, 10000 Zagreb, Croatia.}

\begin{abstract}
The possibility of cluster emission from trans-lead (86$\leq$Z$\leq$96) region of periodic chart has been explored comprehensively by employing few empirical formulas which are modified by adding angular momentum ($l$) or isospin-dependent ($I=(N-Z)/A$) or both terms for the calculation of cluster decay half-lives. These modified versions of the formulas are found with lesser ${\chi}^2$ per degree of freedom and root mean-square error, in addition to the smaller values of some other statistical parameters, while compared to their corresponding old versions on available 61 experimental data of cluster radioactivity. By applying the modified version of the formula given by Balasubramaniam \textit{et al.} [PRC 70 (2004) 017301], the most accurate formula among these, half-lives of several clusters i.e. isotopes of Be, B, C, N, O, F, Ne, Na, Mg, and Si are predicted systematically for the several isotopes in the trans-lead region. The contest of cluster emission with $\alpha$-decay has been investigated in form of branching ratio which brings several potential cluster emissions into the probable decay modes of these nuclei. The accurate prediction of half-lives of such clusters is expected to be crucial for the future experimental observations where $\alpha$-decay is observed dominantly.
\end{abstract}




\end{frontmatter}


\section{Introduction}
\label{intro}
In 1980, Sandulescu \textit{et al.} \cite{sandulescu1980} firstly predicted a new type of radioactivity: cluster radioactivity, which was based on fragmentation theory, where fusion and fission reaction valleys were generated by the shell closure effect \cite{Gupta1994}. Later in 1984, Rose and Jones experimentally proved the existence of this new type of exotic decay \cite{rose1984}, in which $^{14}$C decays from actinide parent nucleus $^{223}$Ra and forms a stable doubly magic (Z=82, N=126) nucleus $^{208}$Pb. Till now, many clusters decays from light to heavy clusters ($^{14}$C to $^{32}$Si) have been observed from various trans-lead nuclei (Fr, Ra, Ac, Pa, Th, U, Pu, etc.) resulting the corresponding daughter nuclei as magic nuclei (Z=82) or neighboring ones (Z=80, 81, and 83), which indicate the importance of shell and pairing effects in cluster radioactivity \cite{Kumar2003,Kumar2009,Gupta2003}. These clusters are observed with long half-lives (T$_{1/2}$) in the range 10$^{11}$-10$^{30}$ sec. \cite{Bonetti2007}.\par
Theoretically, the half-lives of cluster emissions are predicted using various models such as unified fission model (UFM) \cite{Shi1989}, generalised liquid drop model (GLDM) \cite{Royer2001}, super-asymmetric fission model (SAFM) \cite{Poenaru1985}, preformation cluster model (PCM) \cite{Poenaru2002}, etc. Cluster decay half-lives are also calculated by using various semi-empirical formulas such as (i) the empirical relation suggested by Balasubramaniam \textit{et al.} (BKAG formula) for cluster decay half-lives with only three parameters \cite{balasubramaniam2004}, (ii) the empirical relation suggested by Ren \textit{et al.} (RenA formula) using a microscopic density-dependent cluster model with the re-normalized M3Y nucleon-nucleon interaction \cite{rena2004}. Concomitantly, based on experimental observations about the characteristics of exotic cluster decays, scaling law proposed by Horoi \cite{horoi2004} in which logarithmic half-life is proportional to scaling variable $(Z_{c}Z_{d})^{0.6}/\sqrt{Q}$ and also proportional to $\sqrt{\mu}$, where $\mu$ is the reduced mass of cluster and daughter nuclei which was followed by another semi-empirical formula (NRDX), proposed by Ni \textit{et al.} \cite{nrdx2008} considering WKB barrier penetration probability with some approximations. In 2009, Qi \textit{et al.} introduced universal decay law (UDL) \cite{udl2009} that originates from the mechanism of charged particle decay and R-matrix for all sort of decays of clusters, which includes monopole radioactive decays as well. Poenaru \textit{et al.} \cite{univ2011} plotted a universal curve (UNIV) which is found to be a straight line for cluster decay and $\alpha$-decay.\par

All the above-mentioned formulas have been fitted to the available experimental data without considering the dependence of half-lives on angular momentum taken away by the cluster: expected to be crucial alike to the $\alpha$-decay \cite{Denisov2009} to delineate all sets of experimental data. The importance of angular momentum on the $\alpha$-decay half-lives has already been established in a few of our recent works \cite{saxenaQF,pksharma2021} which has invoked us to probe similar dependence on the cluster decay half-lives. In addition to this, isospin ($I=(N-Z)/A$) of parent nucleus is found to be pivotal for the case of $\alpha$-decay in heavy and superheavy nuclei \cite{pksharma2021,saxenaJPG,Singh2021,Soylu2018,Akrawy2019,akrawy2022EPJA} pointing towards its significance in terms of cluster decay as well. Considering these two effects together, modified UDL formula (new UDL) by Soylu and Qi \cite{Soylu2021}, and improved NRDX formula (named as improved unified formula (IUF)) by Ismail \textit{et al.} \cite{Ismail2022} have explained recently that angular momentum and isospin are indeed crucial quantities in determining the cluster decay half-lives.
Importance of isospin effect is also probed by improving semi-empirical formula (ISEM) for the cluster radioactivity in Ref. \cite{cheng2022}. \par

In this article, we have modified the BKAG \cite{balasubramaniam2004}, RenA \cite{rena2004}, Horoi \cite{horoi2004}, NRDX \cite{nrdx2008}, UDL \cite{udl2009}, and UNIV \cite{univ2011} formulas by investigating the effect of centrifugal barrier and isospin terms. These six modified formulas are fitted by using 61 experimental cluster decay data \cite{Bonetti2007,Price1989,Royer2001,Soylu2021}. The comparison of RMSE (root mean square error) between the older and modified version manifestly shows the significance of inclusion of angular momentum and isospin-dependent terms in cluster emission. Furthermore, one of the modified formulas i.e. MBKAG formula (emerged with least RMSE) is employed to calculate the cluster decay half-lives for various cluster emissions like isotopes of Be, B, C, N, O, F, Ne, Na, Mg, and Si in trans-lead region (86$\leq$Z$\leq$96). For these theoretical estimates, the requirement of disintegration energy ($Q$-value) is tested by 121 available experimental $Q$-values \cite{Bonetti2007,Price1989,Royer2001,Soylu2021} from  various mass models \cite{ws42014,moller2019,Singh2012,hfb2004}. Consequently, various potential clusters are proposed from trans-lead region along with their accurate estimation of half-lives. \par
\section{Formalism}
In 2004, Balasubramaniam \textit{et al.} fitted a formula (BKAG) \cite{balasubramaniam2004} for cluster decay. In the course of that year, Ren \textit{et al.} established a formula \cite{rena2004} that can be treated as a natural extension of the Geiger-Nuttall law \cite{geigernuttall1991} as well as the Viola-Seaborg formula \cite{viola1966} from simple $\alpha$-decay to complex cluster radioactivity. In the same year, Horoi also suggested an independent model for $\alpha$-decay which was generalized for cluster emission \cite{horoi2004}. In 2008, Ni \textit{et al.} established NRDX semi-empirical formula for the calculation of half-lives of $\alpha$ and cluster decays \cite{nrdx2008}. Afterwards, Qi \textit{et al.} has introduced universal decay law (UDL) \cite{udl2009} which is widely used by many authors for the estimation of half-lives of cluster radioactivity. In 2011, Poenaru \textit{et al.} fitted UNIV formula \cite{univ2011} and represented a single line of the universal curve on the graph for $\alpha$-decay and cluster decay. The original versions of these formulas are mentioned below:
\begin{eqnarray}
 log_{10}T_{1/2}^{BKAG}(sec.) = [aA_{c}(A_{d}-A_{c})/A + bZ_{c}(Z_{d}-Z_{c})/Z]Q^{-1/2} + c
 \label{eqbkag}
 \end{eqnarray}
 \begin{eqnarray}
 log_{10}T_{1/2}^{RenA}(sec.) = aZ_{d}Z_{c}Q^{-1/2} + bZ_{d}Z_{c} + c \label{eqrena}
 \end{eqnarray}
\begin{eqnarray}
log_{10}T_{1/2}^{Horoi}(sec.) = (a\sqrt{\mu} + b)[(Z_{c}Z_{d})^{0.607}Q^{-1/2} - 7] + (c\sqrt{\mu} + d)
\label{eqhoroi}
\end{eqnarray}

 \begin{eqnarray}
    log_{10}T_{1/2}^{NRDX}(sec.) = aZ_{c}Z_{d}\sqrt{\frac{\mu}{Q}} + b\sqrt{\mu}(Z_{c}Z_{d})^{1/2} + c
     \label{eqnrdx}
     \end{eqnarray}
    \begin{eqnarray}
    log_{10}T_{1/2}^{UDL}(sec.) &=& aZ_{c}Z_{d}\sqrt{\frac{\mu}{Q}}+b[\mu Z_{c}Z_{d}({A_{c}}^{1/3} + {A_{d}}^{1/3})]^{1/2}+c
    \label{udl}
    \end{eqnarray}
     \begin{eqnarray}
     log_{10}T_{1/2}^{UNIV}(sec.) &=& -log P+log_{10}S -[log_{10}(ln2)-log_{10}{\upsilon}]
    \label{univ}
    \end{eqnarray}
  In the above-mentioned formulas $A_{d}$, $A_{c}$ and $Z_{d}$, $Z_{c}$ denote the mass numbers and atomic numbers of the daughter nucleus and cluster, respectively. $Q$ (in MeV) is the energy released in cluster decay, and $\mu = A_{d}A_{c}/(A_{d}+A_{c})$ is the reduced mass. In Eqn. (\ref{univ}), $-log P$ is determined by $a(\mu  Z_{c}Z_{d} R_{b})^{1/2}[arccos\sqrt{r} - \sqrt{r(1-r)} ], r = R_{a}/R_{b}$ with $R_{a} = 1.2249( {A_{c}}^{1/3}+{A_{d}}^{1/3})$ fm, $R_{b} = 1.43998Z_{d}Z_{c}/Q$ fm, and the logarithmic form of preformation factor is given by $log_{10}S = -b(A_{c}-1)$ along with $[log_{10}(ln2)-log_{10}{\upsilon}]$ = d is the additive constant. The values of fitting coefficients a, b, c, and d of the above mentioned formulas can be found in their respective Refs. \cite{balasubramaniam2004,rena2004,horoi2004,nrdx2008,udl2009,univ2011}. \par
  On account of the importance of angular momentum ($l$) as mentioned above, in the present work, as the first step we have modified these formulas by adding only $l$ dependent term ($l(l+1)$), where $l$ is the minimum angular momentum of cluster particle, which is obtained by following selection rules:
\begin{equation}
   l=\left\{
    \begin{array}{ll}
       \triangle_j\,\,\,\,\,\,
       &\mbox{for even}\,\,\triangle_j\,\,\mbox{and}\,\,\pi_{i} = \pi_{f}\\
       \triangle_{j}+1\,\,\,\,\,\,
       &\mbox{for even}\,\,\triangle_j\,\,\mbox{and}\,\,\pi_{i} \neq \pi_{f}\\
       \triangle_{j}\,\,\,\,\,\,
       &\mbox{for odd}\,\,\triangle_j\,\,\mbox{and}\,\,\pi_{i} \neq \pi_{f}\\
       \triangle_{j}+1\,\,\,\,\,\,
       &\mbox{for odd}\,\,\triangle_j\,\,\mbox{and}\,\,\pi_{i} = \pi_{f}\\
      \end{array}\right.
      \label{lmin}
\end{equation}
here, $\triangle_j$ = $|j_p - j_d - j_c|$ with j$_{p}$, $\pi_{i}$, are the spin and parity values of the parent nucleus, respectively. j$_{d}$ is the spin of the daughter nucleus. $\pi_{f} = (\pi_{d})(\pi_{c})$, in which, $\pi_{d}$ and $\pi_{c}$ are the parities of the daughter nucleus and cluster, respectively. For the purpose of fitting, the data of spin and parity are taken from NUBASE2020 \cite{audi20201}. In the next step, the formulas are also modified by adding isospin $I (=(N-Z)/A)$ dependent term ($I(I+1)$). The accuracy and need of addition of different terms belong to the modified formulas are checked by ${\chi}^2$ per degree of freedom (${\chi}^2$) and RMSE values for various versions, which are listed in Table \ref{RMSE} and calculated by using the following relations:
      \begin{equation}
\chi^2 = \frac{1}{N_{nucl}-N_{p}}\sum^{N_{nucl}}_{i=1}\left(log\frac{T^i_{Th.}}{T^i_{Exp.}}\right)^2
\label{kai}
\end{equation}
   \begin{equation}
\text{RMSE}  = \sqrt{\frac{1}{N_{nucl}}\sum^{N_{nucl}}_{i=1}\left(log\frac{T^i_{Th.}}{T^i_{Exp.}}\right)^2}
\label{rmse}
\end{equation}
where, $N_{nucl}$ is the total number of nuclei (data) and $N_{p}$ is the number of degree of freedom (or no. of coefficients). $T^i_{Exp.}$ and
$T^i_{Th.}$ are the experimental and theoretical values of half-lives for $i^{th}$ data point, respectively.

  \begin{table}[!htbp]
\caption{The ${\chi}^2$ and RMSE of various versions of BKAG, RenA, Horoi, NRDX, UDL, and UNIV formulas for 61 cluster decay data.}
\centering
\def\arraystretch{1.0}
\resizebox{1.0\textwidth}{!}{%
{\begin{tabular}{c|ccccccccccccccccc}
 \hline
\multicolumn{1}{c|}{Formula}&
\multicolumn{2}{c}{BKAG}&
\multicolumn{1}{c}{}&
\multicolumn{2}{c}{RenA}&
\multicolumn{1}{c}{}&
 \multicolumn{2}{c}{Horoi}&
 \multicolumn{1}{c}{}&

\multicolumn{2}{c}{NRDX}&
\multicolumn{1}{c}{}&
\multicolumn{2}{c}{UDL}&
\multicolumn{1}{c}{}&
\multicolumn{2}{c}{UNIV}\\
\cline{2-3}\cline{5-6}\cline{8-9}\cline{11-12}\cline{14-15}\cline{17-18}
\multicolumn{1}{c|}{}&
\multicolumn{1}{c}{${\chi}^2$}&
\multicolumn{1}{c}{RMSE}&
\multicolumn{1}{c}{}&
\multicolumn{1}{c}{${\chi}^2$}&
\multicolumn{1}{c}{RMSE}&
\multicolumn{1}{c}{}&
\multicolumn{1}{c}{${\chi}^2$}&
\multicolumn{1}{c}{RMSE}&
\multicolumn{1}{c}{}&
\multicolumn{1}{c}{${\chi}^2$}&
\multicolumn{1}{c}{RMSE}&
\multicolumn{1}{c}{}&
\multicolumn{1}{c}{${\chi}^2$}&
\multicolumn{1}{c}{RMSE}&
\multicolumn{1}{c}{}&
\multicolumn{1}{c}{${\chi}^2$}&
\multicolumn{1}{c}{RMSE}\\
 \hline
Original             &1.01 & 0.98 && 1.10 & 0.95 && 1.45 & 1.16 &&  0.85 & 0.90 && 1.88 & 1.34 &&0.87& 0.91  \\
With $l$ term only   &0.66 & 0.78 && 0.92 & 0.93 && 0.76 & 0.84 &&  0.66 & 0.78 && 0.51 & 0.69 &&0.65& 0.78  \\
With $l$ and I terms &0.44 & 0.63 && 0.68 & 0.79 && 0.77 & 0.83 &&  0.66 & 0.77 && 0.49 & 0.67 &&0.67& 0.77  \\

\hline
\end{tabular}}}
\label{RMSE}
\end{table}
The investigation of addition of different terms leads to the following conclusion from Table \ref{RMSE}: (i) the addition of $l$-dependent term which reflects the hindrance effect of centrifugal barrier, significantly reduces ${\chi}^2$ and RMSE for all the considered six formulas, (ii) whereas, the addition of $I$-dependent term minimises ${\chi}^2$ and RMSE values only for BKAG and RenA formulas. As a result, the final versions of these modified formulas adopted in the present article are given by:
\begin{eqnarray}
 log_{10}T_{1/2}^{MBKAG}(sec.) = [aA_{c}(A_{d}-A_{c})/A + bZ_{c}(Z_{d}-Z_{c})/Z]Q^{-1/2} + cl(l+1)+ dI(I+1) + e
 \label{eqmbakg}
 \end{eqnarray}
 \begin{eqnarray}
 log_{10}T_{1/2}^{MRenA}(sec.) = aZ_{d}Z_{c}Q^{-1/2} + bZ_{d}Z_{c} + cl(l+1) + dI(I+1) + e \label{eqmrena}
 \end{eqnarray}
\begin{eqnarray}
log_{10}T_{1/2}^{MHoroi}(sec.) = (a\sqrt{\mu} + b)[(Z_{c}Z_{d})^{0.607}Q^{-1/2} - 7] + (c\sqrt{\mu} + d) + el(l+1)
\label{eqmhoroi}
\end{eqnarray}

    \begin{eqnarray}
    log_{10}T_{1/2}^{MNRDX}(sec.) &=& aZ_{c}Z_{d}\sqrt{\frac{\mu}{Q}} + b\sqrt{\mu}(Z_{c}Z_{d})^{1/2} + cl(l+1) + d
     \label{eqmnrdx}
     \end{eqnarray}
    \begin{eqnarray}
    log_{10}T_{1/2}^{MUDL}(sec.) &=& aZ_{c}Z_{d}\sqrt{\frac{\mu}{Q}}+b[\mu Z_{c}Z_{d}({A_{c}}^{1/3} + {A_{d}}^{1/3})]^{1/2}+cl(l+1) + d
\label{mudl}
\end{eqnarray}
   \begin{eqnarray}
     log_{10}T_{1/2}^{MUNIV}(sec.) &=& -logP - log_{10}S + cl(l+1) +d
     \label{equniv}
     \end{eqnarray}
      The coefficients a, b, c, d, and e of these modified formulas are mentioned in Table \ref{coefficients}.
 \begin{table}[!htbp]
\caption{The coefficients of MBKAG, MRenA, MHoroi, MNRDX, MUDL, and MUNIV formulas proposed in the present work.}
\centering
\def\arraystretch{0.6}
\resizebox{0.6\textwidth}{!}{%
{\begin{tabular}{c|ccccccccc}
 \hline
\multicolumn{1}{c|}{Formula}&
\multicolumn{1}{c}{$a$}&
\multicolumn{1}{c}{}&
 \multicolumn{1}{c}{$b$}&
 \multicolumn{1}{c}{}&
\multicolumn{1}{c}{$c$}&
\multicolumn{1}{c}{}&
\multicolumn{1}{c}{$d$}&
\multicolumn{1}{c}{}&
\multicolumn{1}{c}{$e$}\\
 \hline
MBKAG & 6.5279 &&89.2684  &&0.0798&&70.0439 &&-100.4122 \\
MRenA & 1.2947 &&-0.0423  &&0.0771&&89.9255 &&-101.5076 \\
MHoroi& 10.1451&&-23.1954 &&4.4835&&-10.9094&&0.0567\\

MNRDX & 0.3590 &&-1.0063  &&0.0634&&-18.8444&&-\\
MUDL  & 0.3564 &&-0.3199  &&0.0737&&-24.8301&&-\\
MUNIV & 0.2369 &&0.6104   &&0.0648&&-23.7267&&-  \\

\hline
\end{tabular}}}
\label{coefficients}
\end{table}

\section{Results and discussions}
\begin{figure}[!htbp]
\centering
\includegraphics[width=0.8\textwidth]{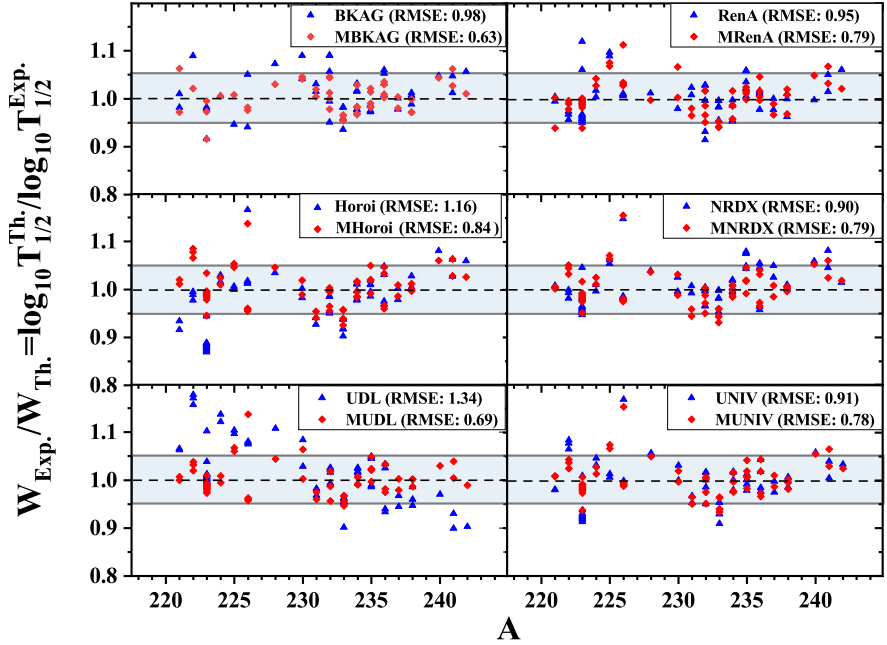}
\caption{(Colour online) Ratio of experimental to theoretical decay widths $W_{Exp.}/W_{Th.}=log_{10}T^{Th.}_{1/2}/log_{10}T^{Exp.}_{1/2}$ for the comparison of our six modified formulas with their respective original versions by using 61 cluster emission data. The RMSE values are also indicated in front of the name of the respective formula.}\label{fig3}
\end{figure}

To ascertain the impact on accuracy for the estimation of half-lives of cluster decay by the addition of the above mentioned terms, we have plotted the ratio of decay widths $W_{Exp.}/W_{Th.}=log_{10}T^{Th.}_{1/2}/log_{10}T^{Exp.}_{1/2}$ as a function of A for our six modified formulas (MBKAG, MRenA, MHoroi, MNRDX, MUDL, and MUNIV) along with their original versions in Fig. \ref{fig3}. Most of the points corresponding to our modified formulas (red diamonds) are between half order of magnitude while the points corresponding to the original formulas (blue triangles) are somewhat widely scattered, which indicate the improvement for the estimation of half-lives of cluster decay after the addition of angular momentum ($l$) or isospin-dependent ($I=(N-Z)/A$) or both terms.

For the comparison among our modified formulas with a few of latest fitted/modified formulas \cite{Soylu2021,Ismail2022,cheng2022} for cluster decay half-lives, we have calculated some other statistical parameters such as standard deviation ($\sigma$), uncertainty ($u$), average deviation factor ($\overline{x}$), and mean deviation $\overline{\delta}$ for 61 experimentally known cluster decay half-lives \cite{Bonetti2007,Price1989,Royer2001,Soylu2021}. All these statistical parameters for these formulas are mentioned in Table \ref{T2}. These statistical parameters are defined as:
\begin{equation}
\sigma = \sqrt{\frac{1}{N_{nucl}-1}\sum^{N_{nucl}}_{i=1}\left(log\frac{T^i_{Th.}}{T^i_{Exp.}}\right)^2}
\end{equation}

\begin{equation}
u =  \sqrt{\frac{1}{N_{nucl}(N_{nucl}-1)}\sum^{N_{nucl}}_{i=1}\left(log\frac{T^i_{Th.}}{T^i_{Exp.}}-\mu \right)^2}
\label{un}
\end{equation}

\begin{equation}
\overline{x} = \frac{1}{N_{nucl}}\sum^{N_{nucl}}_{i=1}\left(\frac{|logT^i_{Exp.}-logT^i_{Th.}|}{logT^i_{Exp.}}\right)
\end{equation}

\begin{equation}
\overline{\delta} = \frac{1}{N_{nucl}}\sum^{N_{nucl}}_{i=1}\left|log\frac{T^i_{Th.}}{T^i_{Exp.}}\right|
\end{equation}
The terms in above equations are already defined in Eqns. (\ref{kai}) and (\ref{rmse}). $\mu$ in Eqn. (\ref{un}) refers to the mean of full data set.

 \begin{table}[!htbp]
\caption{Comparison of MBKAG, MRenA, MHoroi, MNRDX, MUDL, and MUNIV formulas with few others formulas.}
\centering
\def\arraystretch{0.6}
\resizebox{0.6\textwidth}{!}{%
\begin{tabular}{l@{\hskip 0.3in}c@{\hskip 0.3in}c@{\hskip 0.3in}c@{\hskip 0.3in}c@{\hskip 0.3in}c}
\hline
\hline
Formula     & $\sigma$& $u$ & $\overline{x}$&$\overline{\delta}$\\
\hline
MBKAG (Present Work)                &	0.64	&0.08&	0.02	&0.51\\
MRenA(Present Work)                 &	0.80	&0.10&	0.02	&0.62\\
MHoroi (Present Work)               &	0.84	&0.11&	0.03	&0.66\\

MNRDX (Present Work)                &	0.79	&0.10&	0.02	&0.60\\
MUDL (Present Work)                 &	0.70	&0.09&	0.03	&0.53\\
MUNIV (Present Work)                &	0.79	&0.10&	0.03	&0.59\\
New UDL \cite{Soylu2021}            & 0.81    &0.10&  0.03    &0.68\\
IUF \cite{Ismail2022}               & 0.84    & 0.11& 0.03    &0.64\\
ISEF \cite{cheng2022}               & 0.93    &0.12&  0.04    &0.76\\

\hline
\end{tabular}}
\label{T2}
\end{table}

It is clear from Table \ref{T2} that the isospin (only for BKAG and RenA) and angular momentum play a crucial role to improve the cluster decay formulas and result in lesser statistical parameters $\sigma$, $u$, $\overline{x}$, and $\overline{\delta}$ for the modified formulas introduced in the present work, as compared with a few of the latest fitted/modified formulas (new UDL, IUF, and ISEF formulas) for the cluster decay. It is to be noted that among all the modified formulas, MBKAG formula renders more accurate half-life while compared through all the statistical parameters. Hence, MBKAG formula can be employed to predict the more precise half-lives of cluster decay and the probable decay emission. With this in view, the possibility of cluster emission from the experimentally known trans-lead (86$\leq$Z$\leq$96) isotopes is probed by considering the daughter nuclei near the proton shell closure i.e., the emission of a cluster is chosen in such a way that the proton number of daughter nucleus $Z_{d}$ is close to 82 (Pb).\par

\begin{table*}[!htbp]
\caption{The calculated logarithmic half-lives using MBKAG formula together with experimental values \cite{Bonetti2007,Price1989,Royer2001,Soylu2021} for cluster decay. The $\alpha$-decay half-lives are calculated by using NMHF formula \cite{pksharma2021}. BR refers for branching ratios calculated by using Eqn. (\ref{eq}). $Q$ and $Q_{\alpha}$ are the disintegration energies for cluster decay and $\alpha$-decay, taken from Refs.\cite{Bonetti2007,Price1989,Royer2001,Soylu2021} and AME2020 \cite{audii20201}, respectively. For the $l$ values, spin and parity of parent, daughter, and cluster nuclei are used from NUBASE2020 \cite{audi20201}.}
\centering
\resizebox{0.97\textwidth}{!}{%
\begin{tabular}{ccccccccccc}
\hline
\multicolumn{1}{c}{Parent}&
\multicolumn{1}{c}{Daughter}&
\multicolumn{1}{c}{Emitted}&
\multicolumn{1}{c}{$Q$}&
\multicolumn{1}{c}{$Q_{\alpha}$}&
\multicolumn{1}{c}{$l$}&
 \multicolumn{3}{c}{log$_{10}$T$_{1/2}$(sec.)}&
 \multicolumn{1}{c}{BR$_{Exp.}$}&
\multicolumn{1}{c}{BR}\\
 \cline{7-9}
 nucleus &nucleus&cluster&(MeV)&(MeV)&&Exp.&MBKAG&NMHF&&\\
 &&&&&&&(Cluster)&($\alpha$)&&\\

 \hline
$^{221}$Fr & $^{207}$Tl & $^{14}$C  &31.28  & 6.46& 3&14.52  & 15.44  & 2.96  &  -11.56 & -12.48 \\
$^{221}$Ra & $^{207}$Pb & $^{14}$C  &32.39  & 6.88& 3&13.39  & 13.01  & 1.74  &  -11.65 & -11.27 \\
$^{222}$Ra & $^{208}$Pb & $^{14}$C  &33.05  & 6.68& 0&11.22  & 11.46  & 2.32  &  -8.90  & -9.14 \\
$^{223}$Ra & $^{209}$Pb & $^{14}$C  &31.85  & 5.98& 4&15.25  & 15.18  & 5.17  &  -10.08 & -10.01 \\
$^{223}$Ac & $^{209}$Bi & $^{14}$C  &33.06  & 6.78& 2&12.60  & 11.54  & 2.38  &  -10.22 & -9.16 \\
$^{223}$Ac & $^{208}$Pb & $^{15}$N  &39.47  & 6.78& 2&14.76  & 14.36  & 2.38  &  -12.38 & -11.98 \\
$^{224}$Ra & $^{210}$Pb & $^{14}$C  &30.54  & 5.79& 0&15.90  & 15.99  & 5.87  &  -10.03 & -10.12 \\
$^{225}$Ac & $^{211}$Bi & $^{14}$C  &30.48  & 5.94& 4&17.16  & 17.30  & 5.70  &  -11.46 & -11.60 \\
$^{226}$Ra & $^{212}$Pb & $^{14}$C  &28.21  & 4.87& 0&21.19  & 20.68  & 10.52 &  -10.67 & -10.16 \\
$^{226}$Th & $^{212}$Po & $^{14}$C  &30.67  & 6.45& 0&15.30  & 15.02  & 3.79  &  -11.51 & -11.24 \\
$^{228}$Th & $^{208}$Pb & $^{20}$O  &44.72  & 5.52& 0&20.72  & 21.34  & 7.82  &  -12.90 & -13.52 \\
$^{230}$Th & $^{206}$Hg & $^{24}$Ne &57.78  & 4.77& 0&24.64  & 25.78  & 11.91 &  -12.73 & -13.87 \\
$^{230}$U  & $^{208}$Pb & $^{22}$Ne &61.40  & 5.99& 0&19.57  & 20.38  & 6.32  &  -13.25 & -14.06 \\
$^{231}$Pa & $^{207}$Tl & $^{24}$Ne &60.42  & 5.15& 1&23.23  & 23.33  & 10.11 &  -13.12 & -13.22 \\
$^{232}$Th & $^{208}$Hg & $^{24}$Ne &55.62  & 4.08& 0&29.20  & 28.56  & 16.63 &  -12.57 & -11.94 \\
$^{232}$Th & $^{206}$Hg & $^{26}$Ne &55.97  & 4.08& 0&29.20  & 29.21  & 16.63 &  -12.57 & -12.59 \\
$^{232}$U  & $^{208}$Pb & $^{24}$Ne &62.31  & 5.41& 0&21.06  & 21.32  & 9.08  &  -11.98 & -12.24 \\
$^{232}$U  & $^{204}$Hg & $^{28}$Mg &74.32  & 5.41& 0&22.26  & 25.01  & 9.08  &  -13.18 & -15.93 \\
$^{233}$U  & $^{209}$Pb & $^{24}$Ne &60.50  & 4.91& 2&24.82  & 23.71  & 11.86 &  -12.96 & -11.85 \\
$^{233}$U  & $^{208}$Pb & $^{25}$Ne &60.75  & 4.91& 2&24.82  & 23.97  & 11.86 &  -12.96 & -12.12 \\
$^{233}$U  & $^{205}$Hg & $^{28}$Mg &74.24  & 4.91& 3&27.59  & 26.38  & 11.86 &  -15.73 & -14.53 \\
$^{234}$U  & $^{210}$Pb & $^{24}$Ne &58.84  & 4.86& 0&25.88  & 25.06  & 12.19 &  -13.69 & -12.87 \\
$^{234}$U  & $^{208}$Pb & $^{26}$Ne &59.47  & 4.86& 0&25.88  & 25.46  & 12.19 &  -13.69 & -13.27 \\
$^{234}$U  & $^{206}$Hg & $^{28}$Mg &74.13  & 4.86& 0&25.14  & 25.86  & 12.19 &  -12.95 & -13.67 \\
$^{235}$U  & $^{211}$Pb & $^{24}$Ne &57.36  & 4.68& 1&27.42  & 26.95  & 13.37 &  -14.05 & -13.58 \\
$^{235}$U  & $^{210}$Pb & $^{25}$Ne &57.83  & 4.68& 3&27.42  & 27.81  & 13.37 &  -14.05 & -14.43 \\
$^{235}$U  & $^{207}$Hg & $^{28}$Mg &72.20  & 4.68& 1&28.09  & 27.81  & 13.37 &  -14.72 & -14.44 \\
$^{235}$U  & $^{206}$Hg & $^{29}$Mg &72.61  & 4.68& 3&28.09  & 28.70  & 13.37 &  -14.72 & -15.32 \\
$^{236}$U  & $^{212}$Pb & $^{24}$Ne &55.96  & 4.57& 0&25.90  & 28.50  & 14.04 &  -11.86 & -14.46 \\
$^{236}$U  & $^{210}$Pb & $^{26}$Ne &56.75  & 4.57& 0&25.90  & 28.73  & 14.04 &  -11.86 & -14.69 \\
$^{236}$U  & $^{208}$Hg & $^{28}$Mg &71.69  & 4.57& 0&27.58  & 28.40  & 14.04 &  -13.54 & -14.36 \\
$^{236}$U  & $^{206}$Hg & $^{30}$Mg &72.51  & 4.57& 0&27.58  & 28.56  & 14.04 &  -13.54 & -14.52 \\
$^{236}$Pu & $^{208}$Pb & $^{28}$Mg &79.67  & 5.87& 0&21.52  & 21.72  & 7.63  &  -13.89 & -14.09 \\
$^{237}$Np & $^{207}$Tl & $^{30}$Mg &75.02  & 4.96& 2&26.93  & 27.03  & 12.09 &  -14.84 & -14.94 \\
$^{238}$Pu & $^{210}$Pb & $^{28}$Mg &75.93  & 5.59& 0&25.70  & 24.98  & 8.98  &  -16.72 & -16.00 \\
$^{238}$Pu & $^{208}$Pb & $^{30}$Mg &77.03  & 5.59& 0&25.70  & 24.97  & 8.98  &  -16.72 & -15.99 \\
$^{238}$Pu & $^{206}$Hg & $^{32}$Si &91.21  & 5.59& 0&25.30  & 25.27  & 8.98  &  -16.32 & -16.28 \\
$^{240}$Pu & $^{206}$Hg & $^{34}$Si &90.95  & 5.26& 0&25.62  & 26.74  & 10.78 &  -14.84 & -15.96 \\
$^{241}$Am & $^{207}$Tl & $^{34}$Si &93.84  & 5.64& 3&25.26  & 25.94  & 9.21  &  -16.05 & -16.73 \\
$^{242}$Cm & $^{208}$Pb & $^{34}$Si &96.53  & 6.22& 0&23.15  & 23.39  & 6.84  &  -16.31 & -16.55 \\
\hline
\end{tabular}}
\label{table-formulas}
\end{table*}

Before predicting possibilities of new cluster decays in trans-lead regions, we first calculate the half-lives of experimentally known cluster decay using the MBKAG formula which are listed in Table \ref{table-formulas}. We have taken only one parent-cluster combination out of 61 experimental data of cluster decay, to compare with $\alpha$-decay half-lives. For the $\alpha$-decay half-lives, we have used the NMHF (new modified Horoi formula) whose accuracy in determining the half-lives has already been demonstrated in Ref. \cite{pksharma2021}. The first, second, and third columns of Table \ref{table-formulas} show the parent, daughter, and cluster nuclei, respectively. Next two columns represent the disintegration energies of cluster decay and $\alpha$-decay taken from Refs. \cite{Bonetti2007,Price1989,Royer2001,Soylu2021} and from AME2020 \cite{audii20201}, respectively. The sixth column lists angular momentum taken away by cluster particle after emission which is calculated by using selection rules explained in the Eqn. (\ref{lmin}). We have calculated logarithmic half-lives of cluster decay (using Eqn. (\ref{eqmbakg})), tabulated them in the eighth column, and compared these results with the experimental results (presented in the seventh column). It is clear from the Table \ref{table-formulas} that calculated half-lives of cluster emission by using the MBKAG formula (present work) are very close to experimental results. Branching ratio (BR) which quantifies comparison between cluster decay to the $\alpha$-decay and is defined as the ratio of $\alpha$-decay half-life (listed in the ninth column) to the cluster decay half-life as below:
\begin{eqnarray}
 BR = log_{10}b_{c} = log_{10}(\lambda_{c}/\lambda_{\alpha}) = log_{10}(T_{\alpha}/T_{c})
 \label{eq}
 \end{eqnarray}
 where, $\lambda_{\alpha}$ and $\lambda_{c}$ are referred as the decay constants of $\alpha$-decay and cluster emission, respectively. The calculated branching ratios are shown in the last column which are indeed close to experimental branching ratios \cite{Bonetti2007,Price1989,Royer2001,Soylu2021} (presented in the second last column). In fact, an excellent match of half-lives of almost all mentioned clusters in Table \ref{table-formulas} validates the pertinence of MBKAG formula. Furthermore, one can note that the experimental cluster decay half-life goes maximum nearly upto 10$^{30}$ sec., therefore, it can be reasoned out that the clusters with a half-life less than 10$^{30}$ sec. seemingly be of experimental interest.\par
\begin{figure}[!htbp]
\centering
\includegraphics[width=1.00\textwidth]{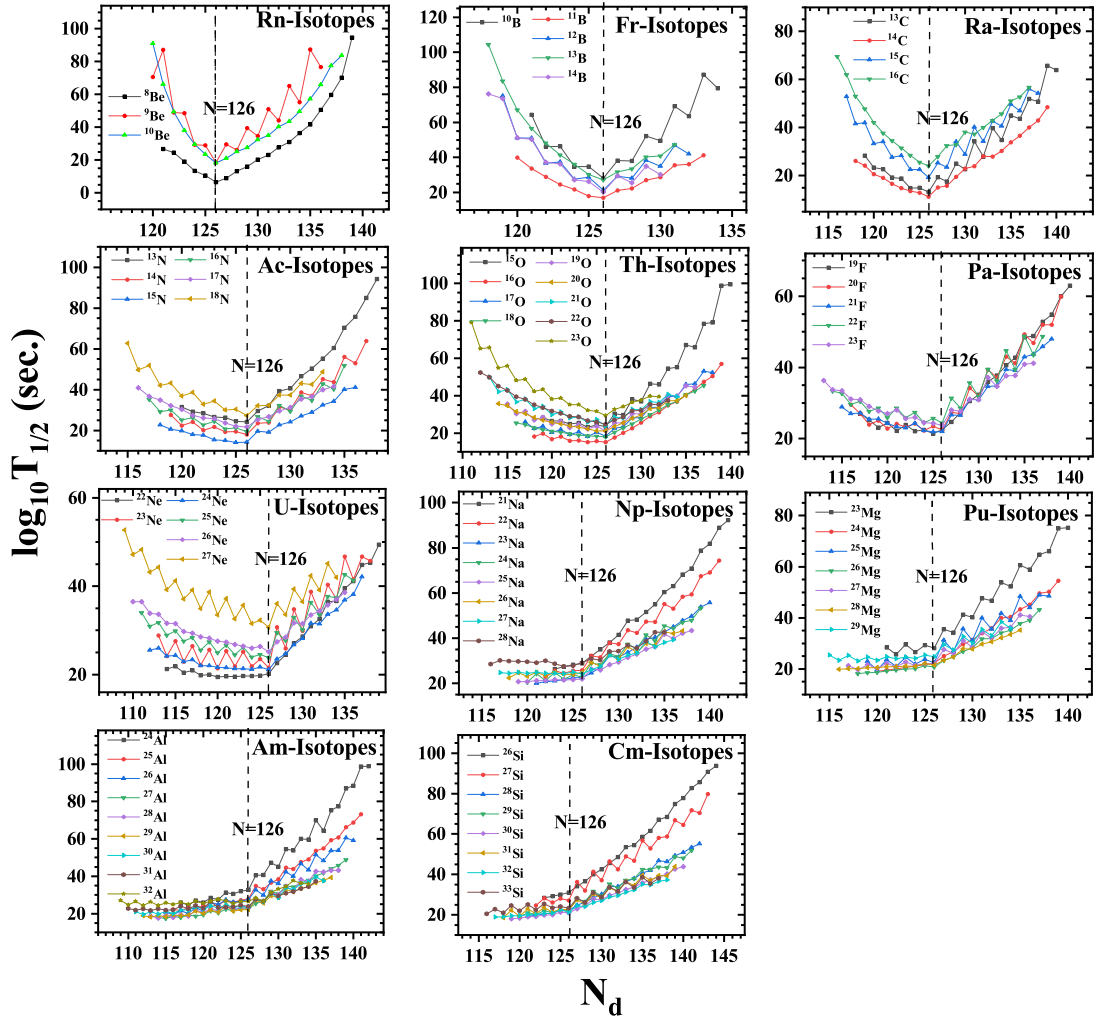}
\caption{(Colour online) Variation of half-lives of various cluster emissions from experimentally known isotopes of trans-lead nuclei (86$\leq$Z$\leq$96) as a function of neutron number of daughter nuclei (considering proton number $Z_{d}$=82). These half-lives are calculated by using MBKAG formula and the $Q$-values are taken from the WS4 mass model\cite{ws42014}.}\label{clusters}
\end{figure}
In the next step of our study, we have utilized the degree of accuracy of MBKAG formula, as exhibited in Table \ref{table-formulas}, to predict the logarithmic half-lives of unknown cluster emissions in the trans-lead region. For this estimation, the $Q$-values are calculated by the following relation:
\begin{eqnarray}
 Q (MeV) = B.E. (d)+ B.E. (c)-B.E. (p) + k[Z_{p}^{\epsilon}-Z_{d}^{\epsilon}]
 \label{Q_value}
 \end{eqnarray}
where, the term $k[Z_{p}^{\epsilon}-Z_{d}^{\epsilon}]$ indicates screening effect caused by the surrounding electrons around the nuclei \cite{Denisov2009prc} with k=8.7 eV [8.7 $\times$ 10$^{-6}$MeV] and $\epsilon$=2.517 for Z (proton number) $\geq$ 60, and k=13.6 eV [13.6 $\times$
10$^{-6}$MeV] and $\epsilon$ =2.408 for Z $<$ 60 have been deducted from the data shown by Huang \textit{et al.} \cite{Huang1976}. For accurate prediction of theoretical $Q$-values, we have selected an effective and reliable possible treatment among various theoretical approaches viz. relativistic mean-field theory (RMF) \cite{Singh2012,saxenaIJMPA2019,saxenaPLB2019,saxena2017,singh2020,saxenaplb2017}, Finite Range Droplet Model (FRDM) \cite{moller2019}, nonrelativistic Skyrme Hartree-Fock-Bogoliubov (HFB) \cite{hfb2004}, and Weizsacker-Skyrme mass model (WS4) \cite{ws42014}. From these approaches, we have calculated RMSE, listed in Table \ref{T3}, for the known 121 $Q$-values related to cluster emissions \cite{Bonetti2007,Price1989,Royer2001,Soylu2021}. Table \ref{T3} establishes that WS4 mass model provides an excellent agreement with the minimum RMSE compared to all other considered theoretical approaches and hence justifies the calculation of $Q$-values for cluster emission by taking binding energies (for daughter(d), cluster(c), and parent(p) nuclei) from this mass model \cite{ws42014}. \par

\begin{table}[!htbp]
\caption{RMSE of various mass models for $Q$-value data for cluster emission.}
\centering
\resizebox{0.5\textwidth}{!}{%
\begin{tabular}{l@{\hskip 2.0in}c}
\hline
\hline
Theory&RMSE\\
\hline
WS4&0.43\\
FRDM&0.78\\
HFB&1.17\\
RMF&3.61\\
\hline
\hline
\end{tabular}}
\label{T3}
\end{table}

After the selection of efficacious empirical formula as well as the theoretical $Q$-values, we have chosen all the parent-cluster combinations for this extensive study to find the possible clusters emitted from $^{211-231}$Rn, $^{213-226}$Fr, $^{214-235}$Ra, $^{215-233}$Ac, $^{216-237}$Th, $^{218-241}$Pa, $^{228-243}$U, $^{226-245}$Np, $^{226-245}$Pu, $^{227-248}$Am, and $^{231-252}$Cm isotopes leading to $^{208}$Pb daughter (doubly magic) and neighbouring nuclei. We have plotted our results (up to T=10$^{100}$ sec.) in Fig. \ref{clusters} where the minima of log$_{10}$T$_{1/2}$ in several panels (Ra-isotopes to U-isotopes) correspond to $^{208}$Pb daughter i.e., doubly magic (Z=82, N=126) or near to it. These minima provide us the most probable clusters emitted from the respective isotopes. However, the probability of cluster emission always competes with $\alpha$-decay which is quantified by branching ratio as we have discussed in Eqn. (\ref{eq}). The limit of experimental branching ratio related to $\alpha$-decay is around $BR=-17$ as can be seen in Table \ref{table-formulas} and also explained by Poenaru \textit{et al.} \cite{Poenaru1991}. Accordingly, cluster emission emerges more probable if $BR\geq-17$: the criteria for the listed probable clusters in Table \ref{Probable}. These clusters are selected from the Fig. \ref{clusters} for the particular isotopic chain of parent trans-lead nuclei $^{211-231}$Rn, $^{213-226}$Fr, $^{214-235}$Ra, $^{215-233}$Ac, $^{216-237}$Th, $^{218-241}$Pa, and $^{228-243}$U. Most of our results are within the experimental reach and also in close match with the recent predictions of Refs. \cite{Santhosh2012,Santhosh2021,Adel2017}.\par

\begin{table*}[!htbp]
\caption{The calculated logarithmic half-lives and branching ratios of probable clusters emitted from various isotopes of trans-lead nuclei (86$\leq$Z$\leq$96). Cluster decay and $\alpha$-decay half-lives are calculated by using MBKAG formula (Eqn. \ref{eqmbakg}) and NMHF formula \cite{pksharma2021}, respectively. Disintegration energies ($Q$-values) for the cluster decay and $\alpha$-decay are taken from WS4 mass model \cite{ws42014} and AME2020 \cite{audii20201}, respectively. For the $l$ values, spin and parity of parent, daughter, and cluster nuclei are used from NUBASE2020 \cite{audi20201}.}
 \centering
 \resizebox{0.8\textwidth}{!}{%
 \begin{tabular}{ccccccccc}
 \hline
\multicolumn{1}{c}{Parent}&
\multicolumn{1}{c}{Daughter}&
 \multicolumn{1}{c}{Emitted}&
\multicolumn{1}{c}{$Q$}&
\multicolumn{1}{c}{$Q_{\alpha}$}&
\multicolumn{1}{c}{$l$}&
 \multicolumn{2}{c}{log$_{10}$T$_{1/2}$(sec.)}&
 \multicolumn{1}{c}{BR}\\
 \cline{7-8}
  nucleus &nucleus&cluster&(MeV)&(MeV)&&MBKAG&NMHF&\\
          &       &       &     &     &&(Cluster)&($\alpha$)&\\
  \hline
$^{216}$Rn& $^{208}$Pb & $^{8 }$Be & 17.13  & 8.20& 0 &6.65  &-2.84 &-9.49   \\
$^{222}$Fr& $^{207}$Pb & $^{14}$B  & 21.56  & 5.85& 0 &20.23 &5.24  &-14.99  \\
$^{221}$Ra& $^{208}$Pb & $^{13}$C  & 31.70  & 6.88& 3 &13.13 &1.74  &-11.39  \\
$^{223}$Ra& $^{208}$Pb & $^{15}$C  & 29.22  & 5.98& 2 &19.15 &5.17  &-13.98  \\
$^{222}$Ac& $^{208}$Pb & $^{14}$N  & 35.64  & 7.14& 1 &17.93 &1.03  &-16.90  \\
$^{222}$Ac& $^{207}$Pb & $^{15}$N  & 39.10  & 7.14& 1 &14.09 &1.03  &-13.06  \\
$^{224}$Ac& $^{208}$Pb & $^{16}$N  & 36.43  & 6.33& 2 &19.44 &3.99  &-15.45  \\
$^{225}$Ac& $^{208}$Pb & $^{17}$N  & 35.64  & 5.94& 2 &21.68 &5.70  &-15.98  \\                                                                                                                       																			
$^{224}$Th& $^{208}$Pb & $^{16}$O  & 46.63  & 7.30& 0 &15.11 &0.81  &-14.30  \\
$^{225}$Th& $^{208}$Pb & $^{17}$O  & 45.02  & 6.92& 2 &18.39 &2.22  &-16.17  \\
$^{226}$Th& $^{208}$Pb & $^{18}$O  & 45.88  & 6.45& 0 &17.98 &3.79  &-14.19  \\
$^{227}$Th& $^{208}$Pb & $^{19}$O  & 44.36  & 6.15& 2 &21.19 &5.16  &-16.03  \\
$^{228}$Th& $^{208}$Pb & $^{20}$O  & 44.87  & 5.52& 0 &21.12 &7.96  &-13.16  \\
$^{229}$Th& $^{208}$Pb & $^{21}$O  & 43.41  & 5.17& 0 &23.84 &9.77  &-14.37  \\
$^{230}$Th& $^{208}$Pb & $^{22}$O  & 43.48  & 4.77& 0 &24.73 &11.91 &-12.82  \\
$^{231}$Th& $^{208}$Pb & $^{23}$O  & 41.08  & 4.21& 2 &29.26 &15.75 &-13.51  \\
$^{228}$Pa& $^{208}$Pb & $^{20}$F  & 50.90  & 6.26& 2 &22.42 &5.13  &-17.29  \\
$^{229}$Pa& $^{208}$Pb & $^{21}$F  & 51.83  & 5.84& 0 &21.94 &6.74  &-15.20  \\
$^{231}$Pa& $^{208}$Pb & $^{23}$F  & 52.01  & 5.15& 1 &23.75 &10.11 &-13.64  \\
$^{231}$U & $^{208}$Pb & $^{23}$Ne & 60.99  & 5.58& 0 &21.55 &8.53  &-13.02  \\
$^{231}$U & $^{206}$Pb & $^{25}$Ne & 59.91  & 5.58& 2 &23.95 &8.53  &-15.42   \\

\hline \end{tabular}}
\label{Probable}
\end{table*}

On the other side, in the panels from Np-isotopes to Cm-isotopes in Fig. \ref{clusters}, in-spite of a clear minima, there is incessantly some probability of emission of clusters since many of the clusters own half-lives less than 10$^{30}$ sec. (experimental limit of half-lives of cluster emissions). For examples, $^{21}$Na from $^{226-229}$Np, $^{22}$Na from $^{226-230}$Np, $^{23}$Na from $^{226-233}$Np, $^{24}$Na from $^{226-234}$Np, $^{25,27}$Na from $^{226-237}$Np, $^{26}$Na from $^{226-236}$Np and $^{28}$Na from $^{224-236}$Np. Similarly, some possible clusters (Mg-isotopes) emitted from various Pu-isotopes (Z$_{p}$=94) are $^{23}$Mg from $^{226-231}$Pu, $^{24,25}$Mg from $^{226-235}$Np, $^{26}$Mg from $^{226-238}$Np, $^{27}$Mg from $^{226-239}$Np, and $^{28,29}$Mg from $^{226-241}$Np. Among Am-isotopes the potential clusters are $^{24}$Al from $^{227-230}$Am, $^{25}$Al from $^{227-233}$Am, $^{26}$Al from $^{227-236}$Am, $^{27}$Al from $^{227-239}$Am, $^{28}$Al from $^{227-240}$Am, $^{29}$Al from $^{227-241}$Am, and $^{30-32}$Al from $^{227-242}$Am as well as $^{26-33}$Si from the $^{231-252}$Cm isotopes. In the emission of odd mass clusters, the odd-even staggering is noticeable in Fig. \ref{clusters} which is usually attributed to the existence of nucleonic pairing correlations \cite{Friedman2009}. The above-mentioned detailed study about favorable clusters having T$_{1/2}<10^{30}$ sec. is expected to be certainly useful for future experimental inputs.
\section{Conclusions}
Several empirical formulas are investigated by adding angular momentum and isospin dependence. Their modified versions are turned into MBKAG, MRenA, MHoroi, MNRDX, MUDL, and MUNIV formulas. Experimental data of a total of 61 nuclei have been utilized for fitting which offers improved results of all the modified formulas while compared to their earlier versions. Among these six modified formulas, after comparison of several statistical parameters the MBKAG formula is found most precise which is used to examine cluster decay half-lives for trans-lead region: $^{211-231}$Rn, $^{213-226}$Fr, $^{214-235}$Ra, $^{215-233}$Ac, $^{216-237}$Th, $^{218-241}$Pa, $^{228-243}$U, $^{226-245}$Np, $^{226-245}$Pu, $^{227-248}$Am, and $^{231-252}$Cm isotopes leading to $^{208}$Pb daughter (doubly magic) and neighbouring nuclei. We have found the considerable probability of emission of various isotopes of Be, B, C, N, O, F, Ne, Na, Mg, and Si from above mentioned trans-lead nuclei, respectively, and many of them are found to be favorable for the measurement (T$_{1/2}<10^{30}$ sec.). This study reveals that doubly magic daughter nuclei play a crucial role in the cluster decay process and could serve as a stimulus to the experiments eyeing on cluster radioactivity.\par
\section{Acknowledgement}
AJ and GS acknowledge the support provided by SERB (DST), Govt. of India under CRG/2019/001851 and SIR/2022/000566, respectively.

\end{document}